# Routing of valley photons in a WS$_2$ monolayer via delocalized Bloch modes of in-plane inversion-symmetry broken photonic crystal slabs

Jiajun Wang[1,*], Han Li[2,*], Yating Ma[2], Maoxiong Zhao[1], Wenzhe Liu[1], Bo Wang[1], Shiwei Wu[1,3], Xiaohan Liu[1,3], Lei Shi[1,3,†], Tian Jiang[2,‡], and Jian Zi[1,3,§]

[1] State Key Laboratory of Surface Physics, Key Laboratory of Micro- and Nano-Photonics Structures (Ministry of Education) and Department of Physics, Fudan University, Shanghai 200433, China

[2] College of Advanced Interdisciplinary Studies, National University of Defense Technology, Changsha 410073, China

[3] Collaborative Innovation Center of Advanced Microstructures, Nanjing University, Nanjing 210093, China

[*] These authors contributed equally to this work.

[†] lshi@fudan.edu.cn

[‡] tjiang@nudt.edu.cn

[§] jzi@fudan.edu.cn

**Abstract**

The valleys of two-dimensional transition metal dichalcogenides (TMDCs) offer a new degree of freedom for information processing. To take advantage of this valley degree of freedom, on one hand, it is feasible to control valleys by utilizing different external stimuli like optical and electric fields. On the other hand, nanostructures are also used to separate the valleys by near field coupling. However, for both above methods, either required low-temperature environment or low degree of coherence properties limit their further applications. Here, we demonstrate all-dielectric photonic crystal (PhC) slabs without in-plane inversion symmetry ($C_2$ symmetry) could separate and route valley photons in a $WS_2$ monolayer at room temperature. Coupling with circularly polarized photonic Bloch modes of such PhC slabs, valley photons emitted by a $WS_2$ monolayer are routed directionally and efficiently separated in the far field. In addition, the far-field emission is directionally enhanced and with long-distance spatial coherence property.

**Introduction**

The emergence of two-dimensional transition metal dichalcogenides (TMDCs) have attracted tremendous interest for their possible applications in valleytronics[1–12]. Due to the broken inversion symmetry in TMDCs, two types of degenerate yet inequivalent valleys (labeled as K and K' valleys) appear at the corners of the first Brillouin zone, shown in Fig. 1a. Interband transitions at valleys, which are excitonic transitions in nature for TMDCs, show highly valley-dependent optical selection rules[4-6, 13–18]. This controllable selective population of certain valleys, called valley polarization, offers a new valley degree of freedom, spawning an emergent field of valleytronics.

To develop valleytronics devices based on TMDCs, effective approaches to separate valleys in the near or far field are indispensable. One feasible way is to selectively excite valleys by utilizing different external stimuli like optical and electric fields[14-18], while the usually required low-temperature environment makes it difficult for the practical applications. Due to the powerful ability of manipulating light, nanostructures[19-21] are also proposed to separate valleys[22–32]. For example, based on either transverse spin momentum of surface plasmons[27, 28] or variable geometric phase of metasurfaces[31], valley separation was achieved in the near or far field at room temperature. However, both the intrinsic loss of metal materials and the localized spatial distribution of resonant modes of nanoantennas limit the efficient valley separation, leading to low degree of valley polarization[24-30]. As a counterpart of metasurfaces, photonic crystals (PhCs) eliminate all these disadvantages due to delocalized photonic Bloch modes and low-intrinsic-loss dielectric constituents. Besides, their Bloch modes are found with peculiar polarization properties. With these attractive properties, PhCs have been widely applied in various researches, such as bound states in the continuum[33-37], topological valley photonics[38, 39], PhC lasers[40-42] and spontaneous emission control of TMDCs[43]. However, to date, there are no reports of effective valley separation in TMDCs by using PhCs.

In this article, we demonstrate that two-dimensional all-dielectric PhC slabs without in-plane inversion symmetry can be used to efficiently separate valley photons in a $WS_2$ monolayer in the far field at room temperature. The valley photons are routed with high directionality and high degree of valley polarization, as is shown in Fig. 1d. For this type of PhC slabs, paired delocalized Bloch modes with different circular polarization not only play a critical role in separating and enhancing directional emission of valley photons, but also lead to spatial coherence properties of the emission field, which are neglected to be discussed in the past studies. Experimentally, the angle-resolved PL results directly show efficient valley separation in the far field with the degree of valley polarization up to 88%. Time-resolved photoluminescence (PL) measurements indicate a 75%-enhanced exciton radiative rate. In addition, the double-slit interference results characterize that the spatial coherence length of emission field on $WS_2$ monolayer is longer than 6 microns (29 microns in theory).

**Principle of separating and routing valley photons**

Analogous to electronic band structures in solids, the Bloch scattering by periodic artificial atoms of PhC slabs alters the dispersion relation of light in the slab, resulting in photonic bands[44]. Each optical state in photonic bands corresponds to the delocalized Bloch mode with well-defined energy and momentum. Modes above the light cone are radiative due to coupling to the free space[44]. For these radiative modes, their polarization states in the far field are strictly defined. And the corresponding polarization states of radiative Bloch modes in an arbitrary photonic band could be further projected into the structure plane and mapped onto the Brillouin zone, defining a polarization field in the momentum space[33, 34]. These polarization properties in principle could be used to control the radiation of luminescent materials. However, owing to high rotation symmetry, the polarization field is nearly linear in most PhC slabs[45]. As a consequence, the polarization states of those PhC slabs could only cover a belt near the equator of the Poincaré sphere[46](a space to describe all polarization states, shown in Fig. 1b). With a large area including two poles not covered, it is useless for us to utilize these Bloch modes of PhC slabs to separate valley photons in TMDCs.

On the contrary, as we know, the broken inversion symmetry is of vital importance in the appearance of inequivalent valley excitons in TMDCs. Similarly, we recently reported that paired circularly polarized states with different chirality would spawn from vortex singularities after breaking the in-plane inversion symmetry of PhC slabs[46], as is shown in Fig. 1c. Then, besides the areas near the equator, the polarization states cover the whole sphere including two poles of the Poincaré sphere, corresponding to polarization states with high degree of circular polarization in momentum space. Therefore, this type of PhC slabs with circularly polarized radiative states could be an ideal platform for us to separate valley photons of TMDCs in the far field, as is illustrated in Fig. 1d.

Firstly, valley photons could couple to circularly-polarized states with corresponding chirality and get separated in the momentum space. Secondly, these Bloch modes are delocalized and could be used in coherent emission[47, 48]. The spatial coherence properties of the emission field lay the foundation for the directional emission and high-efficient separation of valley photons in a $WS_2$ monolayer. Detailed discussion is provided in Supplementary Material section 1.

**Results and discussion**

To demonstrate the existences of the opposite circularly polarized states in momentum space, we designed an in-plane inversion-symmetry broken PhC slab and studied the transmittance spectra in theory and experiment, shown in Fig. 2. The slabs here are made of silicon nitride ($Si_3N_4$, refractive index ~ 2) and silicon dioxide ($SiO_2$, refractive index ~ 1.5).

The thickness of Si$_3$N$_4$ layer is 150 nm. And the thickness of SiO$_2$ layer is 500 microns, which could be considered infinite compared to the wavelength of visible light. Square lattices of holes with a period a = 390 nm are etched in the Si$_3$N$_4$ layer. To break the in-plane inversion symmetry, the shape of the etched hole in a unit cell is set as an isosceles triangle, with the height h and the baseline length b of the triangle being equal (h = d = 250 nm), shown in Fig. 2a. More details about sample design could be found in Supplementary Material section 3.

We first simulated the angle-resolved transmittance spectra under $\sigma_+$ polarized incidence by Rigorous Coupled Wave Analysis (RCWA), with the incidence plane along Γ-X direction. The spectra are asymmetric and there are some diminished regions on the photonic bands, pointed out by blue arrows in Fig. 2b. These diminished regions correspond to the nonexcited states under $\sigma_+$ polarized incidence. Hence those states in the diminished regions are $\sigma_-$ polarized. Changing the incident light to $\sigma_-$ polarization, the diminished regions switch to the other side (Fig. S1b). To show it experimentally, we fabricated samples using electron-beam lithography and reactive ion etching (for more details see Methods). By using homemade polarization-resolved momentum-space imaging spectroscopy system (Fig. S4), angle-resolved transmittance spectra are measured (Fig. 2c), in accordance with the simulation. Both simulated and experimentally measured results confirmed the appearance of optical modes with high degree of circular polarization in our designed PhC slab. For comparison, we also researched the angle-resolved transmittance spectra of the PhC slab with in-plane inversion symmetry. Shown in Fig. 2d, the designed shape of etched hole in the unit is a circle (diameter d = 210 nm). As expected, we did not observe the asymmetric spectra under $\sigma_+$ polarized incidence both in simulation and experiment, shown in Fig. 2e-f. When changing the incidence to $\sigma_-$ polarization, the transmittance spectra is the same as the case of $\sigma_+$ polarization (Fig. S1c). The results demonstrate that by breaking in-plane inversion symmetry of PhC slabs, circularly polarized states would emerge in photonic bands.

The large area of WS$_2$ monolayer is grown on Si/SiO$_2$ substrate by CVD process and then transferred onto PhC slabs. Both PhC slabs and part of unstructured flat Si$_3$N$_4$ substrate are covered (Fig. S10). To study the PL distribution in the far field, angle-resolved PL spectra are measured (Supplementary Material section 5), shown in Fig. 3a-f. The detection plane is along Γ-X direction, in accordance with transmittance spectra measurement in Fig. 2. We selected the $\sigma_+$ ($\sigma_-$) PL by placing a quarter-wave plate and a linear polarizer in the detection path (Fig. S4). Fig. 3e-f show the asymmetric $\sigma_+$ ($\sigma_-$) PL spectra of WS$_2$ monolayer on the PhC slab without in-plane inversion symmetry. The $\sigma_+$ ($\sigma_-$) PL enhanced regions correspond to regions with high degree of $\sigma_+$ ($\sigma_-$) polarization in photonic bands. Fig. 3a-b show $\sigma_+$ ($\sigma_-$) PL spectra of WS$_2$ monolayer on a flat substrate. And Fig. 3c-d show $\sigma_+$ ($\sigma_-$) PL spectra of WS$_2$ monolayer on the PhC slab with in-plane inversion symmetry. Different from Fig. 3e-f, all spectra in Fig. 3a-d are symmetric for both $\sigma_+$ and $\sigma_-$ detection. From the experimental results above, we can draw the conclusion that, as shown in the asymmetric spectra, valley photons emitted by WS$_2$ monolayer have been separated in the far field by PhC slabs without in-plane inversion symmetry. Besides, we performed time-resolved PL measurement at room temperature (Supplementary Material section 10). Compared with WS$_2$ monolayer on a flat substrate, the exciton radiative rate, namely reciprocal of radiative lifetime, is enhanced 75% when WS$_2$ monolayer is on PhC slab without in-plane inversion symmetry.

To further study the degree of separation in Fig. 3e-f, we plotted the angle-resolved $\sigma_+$ ($\sigma_-$)

PL spectra for a single wavelength, shown in Fig. 3g-h. Dot line refers to 615 nm and solid line refers to 628 nm, which are also marked in Fig. 3e-f. We observed that σ+ (red) and σ- (blue) PL maximums separately appear in different angles. The σ+ and σ- PL peaks separate nearly 6 degrees at 615 nm and 3 degrees at 628 nm. For comparison, PL spectra on PhC with in-plane inversion symmetry for corresponding wavelengths are shown in Fig. S5, with the σ+ and σ- PL maximums overlapping in the same angle. We also exhibit the photoluminescence of $WS_2$ monolayer on this PhC slab without in-plane inversion symmetry is highly directional. Shown in Fig. 3g-h, the full width at half maximum of PL peaks (Δθ) is less than 3 degrees at 615 nm and 2 degrees at 628 nm. And this is due to the delocalized property of Bloch modes, leading to long-distance spatial coherence property of the far-field emission by $WS_2$ monolayer on PhC slabs. According to the Fourier relation between momentum and position, wide distribution in the real space means that the mode is localized inside a small area in the momentum space. This corresponds to the small angle distribution of the far-field emission, i.e. the directional emission, and will be further discussed later in this article. For this reason, although the separation of σ+ and σ- PL peaks is small, the valley photons could still be efficiently separated in the far field. Further, we quantify the degree of valley polarization by

$$P(\theta) = \frac{I_+(\theta) - I_-(\theta)}{I_+(\theta) + I_-(\theta)}$$

,where $I_+$ ($I_-$) refers to the PL intensity with σ+ (σ-) polarization for a single wavelength, and θ is the radiation angle. The degree of valley polarization is plotted in Fig. S7, with maximum degree of valley polarization calculated up to 84%. These results indicate the PL of $WS_2$ monolayer on PhC slab without in-plane inversion symmetry is highly directional and with high degree of valley polarization.

Basing on the measured angle-resolved σ+ (σ-) PL spectra of $WS_2$ monolayer on PhC slab without in-plane inversion symmetry, we mapped the PL intensity distribution of a single wavelength in momentum space, shown in Fig. 4a-d. The upper (lower) row corresponds to 615 (628) nm. The PL spectra along different directions in momentum space were measured by rotating the sample in plane relative to the entrance slit of the imaging spectrometer. And the projected momentum k is calculated by $k = k_0 \sin\theta$ ( $k_0 = 2\pi/\lambda$ is the wavevector of light in the free space, θ is the emission angle relative to normal of sample plane). The intensity distribution of σ+ (σ-) PL in momentum space confirmed that the PhC slab without in-plane inversion symmetry leads to directional emission of valley photons.

Then, we used $P(k)$ to qualify the degree of valley polarization in momentum space, which is similarly defined by $P(k) = \frac{I_+(k) - I_-(k)}{I_+(k) + I_-(k)}$, shown in fig. 4e-f. Here, $I_+$ ($I_-$) refers to the PL intensity with σ+ (σ-) polarization for a single wavelength. Experimentally, the maximum $P$ calculated is up to 88%, obtained in Fig. 4f. Note that, the maximum $P$ did not appear along the Γ-X direction in momentum space. The result is as expected for the circular polarized states of the designed PhC slab without in-plane inversion symmetry are slightly shifted from the Γ-X direction in momentum space[40]. The sign of $P(k)$ reverses at opposite sides of momentum space, demonstrating the separation of valley photons with different chirality. In contrast, we also measured and calculated $P(k)$ of the emission by $WS_2$ monolayer placed on a flat substrate, with $P(k)$ being negligible (Fig. S12).

In addition to valley-related directional emission in momentum space, we have expected the spatial coherence property of emission by WS$_2$ monolayer on PhC slab without in-plane inversion symmetry. The Young's double-slit experiments were performed, shown in Fig. 5. The experimental setup is illustrated in Fig. 5a, and the working principle is based on Fourier transformation. The double slit is mounted on the real image plane inside the optical measurement setup to select radiation fields from two different positions on the sample. The radiation fields from these two positions intersect with each other on the Fourier image 2 at the entrance of the spectrometer. Therefore, the spatial coherence properties on the surface of the sample could be directly detected in the far field. Changing the etched depth of PhC slab, we were able to overlap the measured photonic band with the PL spectra of WS$_2$ monolayer, so that we could obtain enough signal intensity. Interference fringes are observed in the angle-resolved PL spectra along Γ-X direction, shown in Fig. 5b. The red-marked line is further plotted in Fig. 5c, showing the interference intensity distribution at 621 nm. The fringe visibility $V$ is calculated around 50%, defined by $V = \frac{I_{max}-I_{min}}{I_{max}+I_{min}}$, where $I_{max}$ and $I_{min}$ are the intensities of adjacent maxima and minima[49]. In this measurement, the real double-slit distance d is 120 microns. The scanning electron microscopy image of the double-slit is presented in Fig. S13. The magnification on the real image is 20, so the effective double-slit distance on the sample is 6 microns. The 6-micron effective double-slit distance is almost ten times the emission wavelength, demonstrating that the measured spatial coherence length is larger than 6 microns. Moreover, the spatial coherence length could be calculated by $\frac{\lambda}{\Delta\theta}$ in theory, which is widely used in the optical coherence theory[50]. Here, $\Delta\theta$ is about 0.0215 (1.23° in degree) at 621 nm (Fig. S14), and the calculated spatial coherence length is around 29 microns. In comparison, no interference fringes are observed when WS$_2$ monolayer is placed on a flat substrate, as shown in Fig. 5b-c. This means that the far-field emission of WS$_2$ monolayer on a flat substrate has no long-distance spatial coherence property. Hence, we reveal that the far-field emission by WS$_2$ monolayer on PhC slab without in-plane inversion symmetry has long-distance spatial coherence property. This property of PhC slab extends the coherence control on PL of WS$_2$ monolayer from temporal coherence to spatial coherence.

In summary, we proposed the in-plane inversion-symmetry broken all-dielectric photonic crystal slabs to route the far-field emission of valley photons of WS$_2$ monolayer at room temperature. By breaking the in-plane inversion symmetry of PhC slab, we observed paired circularly polarized states with different chirality spawn from vortex singularities. Via coupling with those delocalized Bloch modes, valley photons emitted by WS$_2$ monolayer were separated in momentum space, and the exciton radiative rate is significantly enhanced. Besides, both the directional emission and the long-distance spatial coherence property benefit the applications of in-plane inversion-symmetry broken PhC slabs to route valley photons. In addition, our method could be extended to manipulate valley photons of other TMDCs monolayers. And the ability of this PhC slabs to transport valley information from the near field to the far field would help to develop photonic devices based on valleytronics.

**Methods**
**Sample fabrication.** *The fabrication of photonic crystal slab.* The samples' structure was two layers of slab, with a thin silicon nitride layer on the silicon dioxide substrate. The silicon dioxide substrate was cut from a 500-microns-thick quartz wafer. Then a silicon nitride layer was grown on silicon dioxide substrate by plasma-enhanced chemical vapor deposition (PECVD). The thickness of the grown silicon nitride layer was nearly 150 nm and the thickness could be tunable by controlling the deposition time. To fabricate the designed structure, the raw sample was spin-coated with a layer of positive electron beam resist (PMMA950K A4) and an additional layer of conductive polymer (AR-PC 5090.02). Then hole array mask pattern was fabricated onto the PMMA layer using electron beam lithography (ZEISS sigma 300). The sample was further processed by reactive ion etching (RIE). Anisotropic etching was achieved by RIE using $CHF_3$ and $O_2$. The patterned PMMA layer acted as a mask and was eventually removed by RIE using $O_2$. The size of every designed structure is approximately 80 × 80 microns.

*Transfer process for $WS_2$ monolayer.* The CVD $WS_2$ monolayer on $Si/SiO_2$ substrate was spin-coated with poly (L-lactic acid) (PLLA) before baking for 5 minutes at 70°C. Afterwards, a PDMS elastomer was placed on top of PLLA film and then torn off. The composite was then attached to a glass slide and put under the microscope on a transfer stage. The PhC slab placed under the glass slide was aligned carefully using microscope and the glass slide was lowered to contact PhC slab. The stage was heated to 70°C to improve the adhesion, then the glass slide was lifted with PDMS on it, leaving $WS_2$ monolayer on the PhC slabs. After dissolving PLLA in dichloromethane, the $WS_2$ monolayer was finally transferred to designed photonic crystal slabs.

**Optical measurements.** *Experimental measurements of time-resolved PL.* Please see the supplementary Material section 4 for the schematics and discussions.

*Measurement setup of the polarization-resolved momentum-space imaging spectroscopy system and double-slit experiment.* Please see the supplementary Material section 5 for the schematics and discussions.

**Simulations.** The transmittance spectra were simulated by Rigorous Coupled Wave Analysis (RCWA). The periodic boundary conditions were applied in x, y direction. The polarization angle was set $\pi/4$ and phase difference was set $\pi/2$ or $3\pi/2$ to get circularly polarized incidence (The polarization angle 0 ($\pi/2$) corresponds to p (s) polarization). The $Si_3N_4$ refractive index was set to 2 and $SiO_2$ refractive index was set to 1.5. All the materials were considered no loss in visible light.

**Acknowledgements**
We thank Dr. Haiwei Yin from Ideaoptics Corporation for helpful discussions. The work was supported by China National Key Basic Research Program (2016YFA0301103, 2016YFA0302000 and 2018YFA0306201) and National Science Foundation of China (11774063, 11727811 and 91750102, 91963212, 11804387, 11802339, 11805276, 61805282, 61801498, and 11902358), and the Youth talent lifting project (Grant No. 17-JCJQ-QT-004). The research of L. S. was further supported by Science and Technology Commission of Shanghai Municipality (19XD1434600,2019SHZDZX01,19DZ2253000).

**Conflict of interest**
The authors declare that they have no conflict of interest.

**Contributions**

L. S, T. J, and J. Z conceived the basic idea for this work. J. W designed the structure, performed the sample fabrications, and carried out the RCWA simulations. Y. M performed the material transfer. H. L, J. W and Y. M performed the TR-PL measurement. J. W and M. Z performed the angle-resolved spectra measurement. J. W and H. L analyzed the experimental data. L. S, T. J and J. Z supervised the search and the development of the manuscript. J. W and L. S write the draft of the manuscript. All the authors contributed to the discussion of the results and writing the manuscript.

**Data availability**

The data that support the findings of this study are available from the authors on reasonable request, see author contributions for specific data sets.

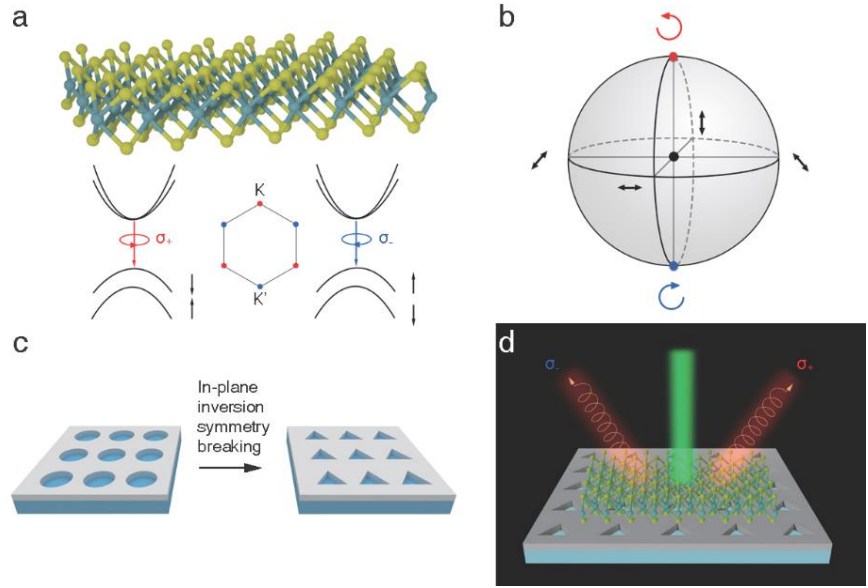

**Fig. 1 Schematics and principle. a,** Schematic of $WS_2$ monolayer and optical selection rules at the K and K' valleys. $\sigma_+$ ($\sigma_-$) excitation corresponds to interband optical transition at the K (K') valley. **b,** The normalized Poincaré sphere. Different azimuthal positions on the equator correspond to different linearly polarized states. Two poles correspond to two types of circularly polarized states. The center of the Poincaré sphere corresponds to vortex singularity. **c,** PhC slabs with $C_4$ symmetry and without in-plane inversion symmetry. **d,** Illustration of photoluminescence of $WS_2$ monolayer on the PhC slab without in-plane inversion symmetry. Valley photons with different chirality separate and radiate directionally in the far field.

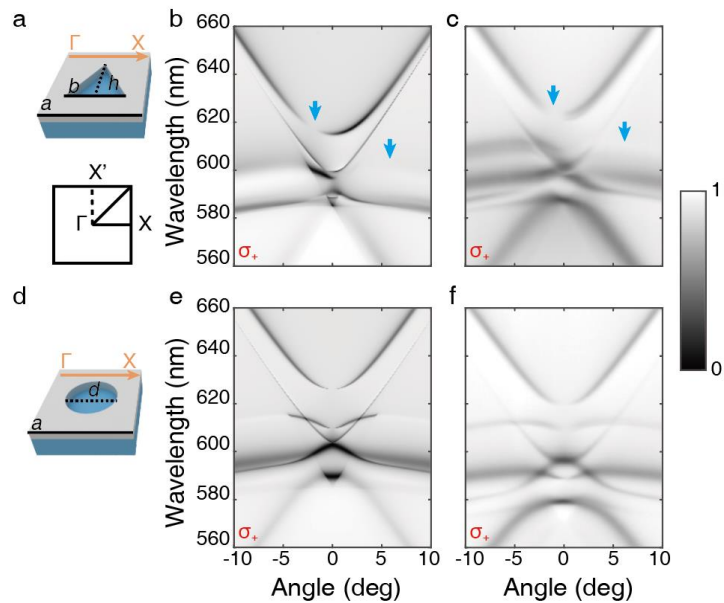

**Fig. 2 Simulated and experimentally measured angle-resolved transmittance spectra. a,** Unit cell of photonic crystal slab without in-plane inversion symmetry. Triangle air holes are etched in the $Si_3N_4$ layer. **b-c,** Simulated and measured angle-resolved transmittance spectra in the visible range under $\sigma_+$ polarized incidence. The incidence plane is along Γ-X direction. Diminished regions pointed out by blue arrows imply the existence of circularly polarized states after breaking the in-plane inversion symmetry. **d,** Unit cell of photonic crystal slabs with in-plane inversion symmetry. Circle air holes are etched in the $Si_3N_4$ layer. **e-f,** Simulated and measured angle-resolved transmittance spectra in the visible range under $\sigma_+$ polarized incidence. The spectra is symmetric for the PhC slab with in-plane inversion symmetry.

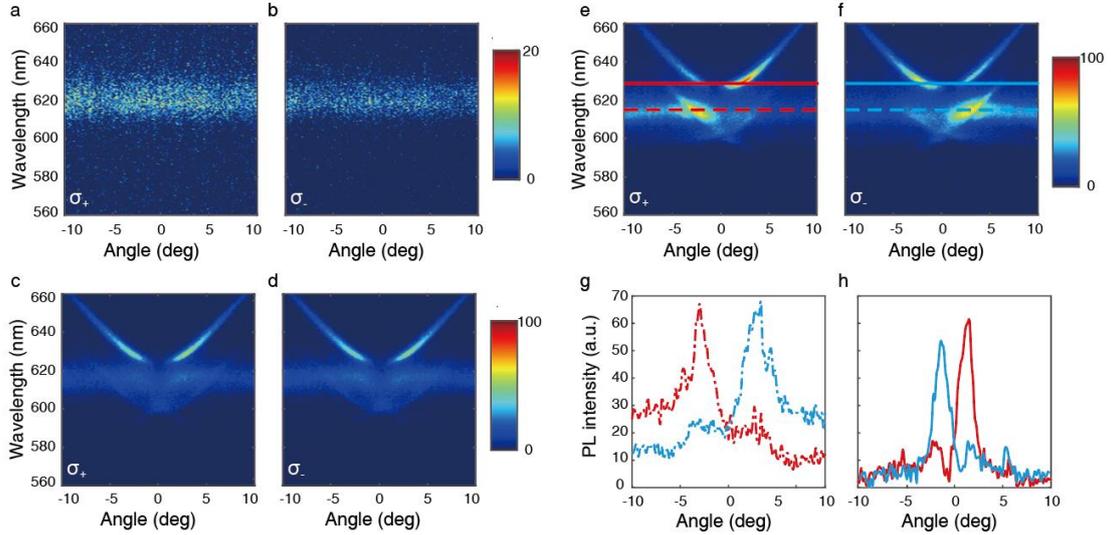

**Fig. 3 Angle-resolved PL spectra of WS$_2$ monolayer on different substrates at room temperature. a-f,** Angle-resolved PL spectra of WS$_2$ monolayer on three different substrates with $\sigma_+$ ($\sigma_-$) detection along Γ-X direction. **a** and **b** correspond to WS$_2$ monolayer on a flat substrate. **c** and **d** correspond to WS$_2$ monolayer on the PhC slab with in-plane inversion symmetry. **e** and **f** correspond to WS$_2$ monolayer on the PhC slab without in-plane inversion symmetry. **g-h,** Separation of $\sigma_+$ (red) and $\sigma_-$ (blue) polarized light at 615 nm (dot line) and 628 nm (solid line) in **e-f**. The maximum $P$ calculated is up to 84%.

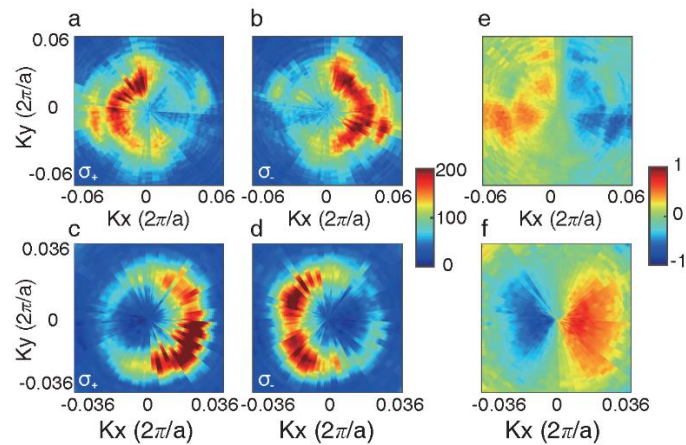

**Fig. 4 Experimental measurement of PL spectra and valley polarization in momentum space. a-d,** $\sigma_+$ and $\sigma_-$ PL intensity distribution in momentum space at 615 nm (upper) and 628 nm (lower). These are for WS$_2$ monolayer on the PhC slab without in-plane inversion symmetry. **e-f,** Images of valley polarization $P(k)$ in momentum space. The maximum $P$ calculated is up to 88%.

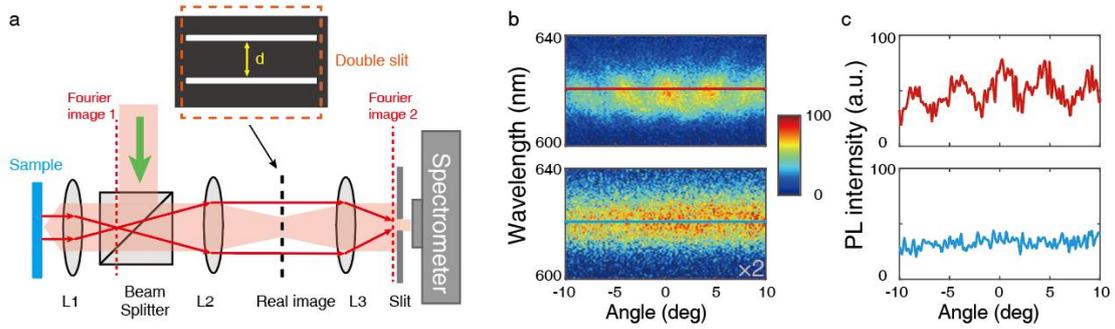

**Fig. 5 Experimental results of Young's double-slit interference. a,** Schematic view of the experimental setup. L, Lens. The double-slit distance is d. The double slit is mounted on the real image plane. The red arrow lines show the radiation fields from two different position on the sample. **b,** The experimental results for the case of a 6-micron effective double-slit distance. The real double-slit distance d is 120 microns. The magnification on the real image is 20, so the effective double-slit distance on the sample is 6 microns. The upper is for $WS_2$ monolayer on PhC slab without in-plane inversion symmetry. The lower is for $WS_2$ monolayer on a flat substrate, with signal intensity shown twice the measurement. The detection plane is along Γ-X direction. **c,** The interference intensity distribution in **b** at 621 nm. The far-field emission of $WS_2$ monolayer on PhC slab without in-plane inversion symmetry has long-distance spatial coherence property.